# ∀ ∃λΣ metamorphoses.

# Generalizations, variations,

# algorithmic semantics

Alex Shkotin,

ashkotin(at)acm.org

**Abstract** This article contains ideas and their elaboration for quantifiers, which appeared after checking in practice the experimental language of the formal knowledge representation YAFOLL [1]:

- looking at ∀ ∃ as operators clarifying two trivial properties of a function: the constancy of result value, presence of a value in the result;

-It turned out that the quantifier term can be written in the lambda calculus technique, i.e. as definition;

-quantifier of quantity # is introduced into the language, as needed in practice and does not cause logical and algorithmic problems on finite structures;

- the quantifier of the sum Σ is mentioned because it is a quantifier of the language;

-algorithmic semantics is written for ∀ ∃ as an introduction to the topic.

**Designations and values in YAFOLL**

Yp - YAFOLL processor. It receives YAFOLL text as an input, tries to execute it and issues various messages.

_True is the truth value of truth.

_False is the truth value of a false.

TV is a sort of truth values. There are only two elements in it: _True, _False.

Number - a sort of rational numbers. Yp perceives them in the usual decimal notation with an optional dot and without an exponential form.

# Introduction

YAFOLL [1] is an experimental formal many-sorted language for defining and maintaining functions on finite sets (sorts) and constructing, with their help, finite models of reality parts, as well as formal theories about the properties of these parts. Its main idea is expressed by the formula:

Ontology = formal theory + finite model.

In the course of experiments to create formal theories and maintain their finite models, some logical relationships and the need to increase the expressive power of the language were clarified.

We further consider some of the quantifiers necessary for ontologies, their possible generalizations, and semantics.

# ∀ ∃ with parameter and two function properties

We generalize a little the use of quantifiers: a sub-quantifier function can have any sort of result, not just TV (a sort of truth values). The sort of value of the quantifier itself is TV.

Let f be a function of signature s:s1, i.e. from sort s to sort s1, and v1 is an element of s1.

f may or may not have the following two properties:

a) the function f is constant on all elements of s with the value v1,

b) the function f on some element s returns the value v1.

Let's consider the nuances and formalization:

(a)

if s is empty then (a) is not satisfied, because v1 is not reached.

if f is partial then (a) is not satisfied, because we need "on all s elements."

Examples and formalization:

"function fn is constant on s with a value of 10."

Formalized ∀(10)x:s fn(x).

"the <u>predicate</u> p is constant on s with the value _True."

Formalized ∀(_True)x:s p(x).

In this small generalization, the quantifier of universality has a parameter - what value is "looking for."

(b)

It is clear that if s is empty then (b) is false.

If f is empty then (b) is false.

Examples and formalization:

"the function fn on some element s has a value of 10."

Formalized ∃(10)x:s fn(x)

"the predicate p on some element s has the value _True."

Formalized ∃(_True)x:s p(x)

In this small generalization, the existential quantifier has a parameter - what value is "looking for."

So it turned out that the quantifiers check if the function has some property.

# Function with external parameters. Transition to terms

In the case when a term can be written in a quantifier expression in place of a function, which is natural for quantifiers, it should be taken into account that in this term can be quantifier variables of upper quantifiers in which the given is embedded, as well as formal parameters if the quantifier is embedded in the term of the function definition. From the point of view of the structure of the term itself, these variables are free in the term and if we look at such a term as a defining function, then it will have external, outside it parameters, similar to the "global", "visible" variables in some programming languages.

Thus, we get a syntactically complete version of using quantifiers with a parameter:

Q (v) x:s t

where t is the usual term for quantifier.

There is also WFC (well-formedness constraint): the sort of v and the sort of value of t must be the same.

# Church quantifier - λ

So, the term of quantifier defines a function of the quantifier's variable, a function in the general case with external parameters.

Defining functions without naming them is the Church lambda calculus, where the meaning of the quantifier λ is "Let's define!", Which is naturally denoted by $\stackrel{def}{=}$ instead of λ.

We write the classic quantifier (the sort of term value under the quantifier is TV)

∀ x:s t

this way

∀ ((s: TV) (x) $\stackrel{def}{=}$ t)

where the quantifier ∀ has turned into an operator whose parenthesis keep the phrase defining a function of signature s:TV, i.e. from s to TV.

In YAFOLL, if we want to define a named function, we write

Declaration f1 (s:TV) definition (x): t.

Where

 (s: TV) definition (x): t

syntactically equivalent

(s: TV) (x) $\stackrel{def}{=}$ t

and it reads: define the function of signature (s:TV) of one variable x, by the term t.

Which, in turn, through λ can be written, for example,

λx: s.t

Thus, the quantifier has in parenthesis the Church's definition of function (in the general case with external parameters), and the quantifier itself turns into an operator over function, figuring out the property of the function passed to it.

Logically and technically, quantifier variables leave our language - they turn into formal parameters.

## # - quantity quantifier

When working with finite structures (for example, graphs or tables of the R-model), obtaining quantitative characteristics is natural and quite possible.

The quantifier of quantity formalizes phrases of the form "the number of elements of the sort x with the property p."

Formally

Let p be a unary predicate.

(#x: s p (x)) returns the number of elements of sort s satisfy the predicate p.

The sort of result is a natural number. In YAFOLL, natural numbers are considered part of the rational sort Number.

If s is empty, 0 is returned.

If p is empty, 0 is returned.

\# is used in the query language about the properties of the domain model, for example, "How many planets in the solar system?".

### Examples

\# is often used in definitions and axioms of the theory of an application area, especially often in requirements for the quantity.

So far, here is an example from mathematics - the theory of undirected graphs:

(1) "an edge with only one end vertex is called a loop."

Let E be the sort of edges, V the sort of vertices, enp the incidence predicate of the vertex and edge. Definition in YAFOLL:

Declaration loop (E:TV) definition (e): (#v: V (v enp e)) = 1.

Where (E:TV) is the signature of the function "loop": from E to TV.

Let the current finite structure with which the Yp processor works contains an undirected graph and _edge1 is an element of sort E.

If Yp got a query ? loop (_edge1)? it will return _True or _False.

(2) An important property of the edges "e is a binary edge" meaning "e is incident to at most two vertices" is formalized as

Declaration E2 (E:TV) definition (e): (#y: V (e inc y)) ≤ 2.

where inc is the incidence predicate of the edge to the vertex.

In many formal theories, the quantifier of quantity is only "syntactic sugar", i.e. may be eliminated, but when working with models, quantitative characteristics are a matter of direct interest.

## Σ - sum quantifier

Let f(x) be the unary function with the signature of s:Number, where s to be the finite sort.

The phrase "the sum of the values of f on elements of sort s" is formalized in YAFOLL as

Σx:s f(x)

the quantifier returns the sum of the values of f on the elements of s.

If s is empty or the function f is empty, Yp will display a "No value" message.

## ∀ ∃. Algorithmic semantics

When we work with finite structures, enumeration of elements of a finite sort does not cause any logical difficulties if we are given an enumeration mechanism: we understand that the enumeration process, once started, will iterate over all elements of the sort and be completed.

Let us look at the property (a) (see the first paragraph) algorithmically.

Statement: "on each element of s, the value of f is equal to v."

**formalization**

(∀(v)x:s f(x))

**algorithm for obtaining the ∀ quantifier value**

If s is empty return _False.

**Iterate** over s having the current value in **x**, performing the following actions with each current value:

if f has no value on x return _False,

if the value of f(x) is not equal to v return _False.

Return _True.

XO

Although the algorithm is written in English, we give some explanations:

- the "iterate" command ends with a dot after the second _False, if the command is completed to the end, we get to the next command after it.

- the algorithm is executed by some processor and "return _False" means to complete the execution of the algorithm with a value _False.

-XO mark indicates the end of the algorithm text.

**Thesis: how to get the ∀ quantifier value is set clearly and correctly:-)**

This is algorithmic semantics:-)

Such semantics can be regularly converted into instructions of a suitable programming language.

Similarly for the property (b) of the first paragraph:

Statement: "on some element of, s the value of f is equal to v."

(∃(v)x:s f(x))

**algorithm for obtaining the ∃ quantifier value**

If s is empty return _False.

**Iterate** over s having the current value in **x**, performing the following actions with each current value:

if f does not have value on x go to the next,

if f(x) is v return _True.

return _False.

XO

Comment. Inside the "iterate" command, a new absolutely understandable subcommand "go to the next" element of s appeared:-)

# Conclusion

We have got a generalization of quantifiers ∀ ∃ leading to Church λ and quantifiers without quantifier variables.

The presence of the quantifier of quantity # is **a watershed between the theoretical and applied languages for ontologies**: in the first quantifier is absent, in the second it is necessary, at least for working with models. Its presence in the formulas of the theory depends on the domain area. Note that OWL2 [2] has phrases in which quantities are used, for example,

ClassAssertion (ObjectExactCardinality (5: hasChild): John)

i.e. the idea of quantity in the language is present.

Algorithmic semantics is attractive in that it is easily programmed and important in that we find out **a set of commands of a logical level** - at this time we have got "iterate" command.

# References

[1] A. Shkotin. Finite Systems Handling Language (YAFOLL message 1), December 2015, Studia Humana 4(4). DOI: 10.1515/sh-2015-0021

https://www.researchgate.net/publication/307841408_Finite_Systems_Handling_Language_YAFOLL_message_1

[2] OWL 2 Web Ontology Language, Primer (Second Edition), W3C Recommendation 11 December 2012. https://www.w3.org/TR/owl-primer/